\documentclass[a4paper,11pt]{article}
\usepackage{jheppub}
\makeatletter
\gdef\@fpheader{ }
\makeatother
\usepackage[utf8]{inputenc}
\usepackage{graphicx,tikz}
\usetikzlibrary{matrix,arrows,decorations.pathmorphing,decorations.markings,calc,shapes}
\usepackage{tikz-cd}
\usetikzlibrary{calc}
\usepackage{amssymb,amsthm,amsmath}
\usepackage{hyperref}

\theoremstyle{plain}

\theoremstyle{definition}

\newcommand{\on}{\operatorname}

\newcommand{\g}{\mathfrak{g}}
\newcommand{\h}{\mathfrak{h}}

\newcommand{\la}{\langle}
\newcommand{\ra}{\rangle}
\newcommand{\R}{\mathbb{R}}
\newcommand{\half}{\frac{1}{2}}

\newcommand{\BE}{\mathsf E}
\newcommand{\CE}{\mathcal E}
\newcommand{\twedge}{{\textstyle\bigwedge}}

\newcommand{\disk}{{\tikz[baseline=-0.8ex]\filldraw[fill=black!10!white] (0,0)circle(0.8ex);}}
\newcommand{\sdisk}{{\tikz[baseline=-0.6ex]\filldraw[fill=black!10!white] (0,0)circle(0.6ex);}}
\newcommand{\annu}{{\begin{tikzpicture}[baseline=-0.8ex]\filldraw[fill=black!10!white] (0,0)circle(0.8ex);\filldraw[fill=white] (0,0)circle(0.3ex);\end{tikzpicture}}}
\newcommand{\sannu}{{\begin{tikzpicture}[baseline=-0.8ex]\filldraw[fill=black!10!white] (0,0)circle(0.8ex);\filldraw[fill=white] (0,0)circle(0.3ex);\end{tikzpicture}}}

\tikzset{->-/.style={decoration={
  markings,
  mark=at position #1 with {\arrow{>}}},postaction={decorate}}}

\title{Poisson-Lie T-duality as a boundary phenomenon of  Chern-Simons theory}
\author{Pavol \v{S}evera}
\affiliation{Section of Mathematics, University of Geneva, Switzerland}
\emailAdd{pavol.severa@gmail.com}

\abstract{%
We give a ``holographic'' explanation of Poisson-Lie T-duality in terms of Chern-Simons theory (or, more generally, in terms of Courant $\sigma$-models) with appropriate boundary conditions. }

\begin{document}
\maketitle

\section{Introduction}
Poisson-Lie T-duality \cite{KS} is a generalization of T-duality, replacing Abelian Lie groups (tori) with non-Abelian Lie groups. As in the Abelian case, it is an equivalence of two (or more) 2-dimensional $\sigma$-models. In the simplest case of no ``spectator coordinates'' it is given by the following data: a Lie group $G$  with an invariant inner product $\la,\ra$ of signature $(n,n)$ on its Lie algebra $\g$, and a vector subspace $\CE_+\subset\g$ of dimension $n$ such that $\la,\ra$ is positive-definite (or at least non-degenerate) on $\CE_+$. 

If  $H\subset G$ is a closed subgroup such that its Lie algebra $\h\subset\g$ is Lagrangian (i.e.\ $\h^\perp=\h$) then this data produces a Riemannian metric and a closed 3-form on $G/H$. If $H'\subset G$ is another Lagrangian subgroup then Poisson-Lie T-duality is an equivalence of the 2-dimensional $\sigma$-models with targets $G/H$ and $G/H'$. (In the case of the ordinary (Abelian) T-duality $G,H,H'$ are tori).

Indeed, the sigma models can be almost entirely described just in terms of $G$ and $\CE_+$, i.e.\ independently of $H$ (or $H'$). This is best seen for the Hamiltonian descriptions of the $\sigma$-models, but let us summarize also some other features:

\subsubsection*{Equations of motion}
If $\Sigma$ is a surface with a pseudo-conformal structure, typically a cylinder, then a map
$$f:\Sigma\to G/H$$
 is a solution of the equations of motion iff $f$ admits a lift 
$$g:\tilde\Sigma\to G$$
 where $\tilde\Sigma$ is the universal cover of $\Sigma$, such that
\begin{equation}\label{eq-motion}
\partial_+g\,g^{-1}\in \CE_+\otimes\Omega^1(\Sigma),\quad \partial_-g\,g^{-1}\in \CE_-\otimes\Omega^1(\Sigma)
\end{equation}
where $\CE_-=(\CE_+)^\perp$ and  $d=\partial_{+}+\partial_-$ is the splitting of $d$ to  the light-like directions. The lift $g$, if it exists, is unique up to  right multiplication by a (constant) element of $H$. If the lift $g$ is actually a map $g:\Sigma\to G$ then we can project $g$ to a solution of the Euler-Lagrange equations $f':\Sigma\to G/H'$. This condition is the \emph{non-Abelian momentum constraint}; in the case of Abelian T-duality (when $G$ is a torus), it is the momentum quantization condition.

Various dynamical quantities can be read off the lift $g$. For example,
the energy-momentum tensor on $\Sigma$ is equal to $\half\la\partial_+g\,g^{-1},\partial_+g\,g^{-1}\ra-\half\la\partial_-g\,g^{-1},\partial_-g\,g^{-1}\ra\in\Gamma(S^2T^*\Sigma)$.

\subsubsection*{Hamiltonian description}
The phase space of the $G/H$ (and also $G/H'$) $\sigma$-model, with the non-Abelian momentum constraint imposed, is $(LG)/G$ with its standard symplectic structure (without the constraint it is the space of maps $g:\R\to G$ such that $g(x+2\pi)=g(x)h$ for some $h\in H$, modulo the right action of $H$ by multiplication). The Hamiltonian is (cf.\ the energy-momentum tensor given above)
$$\mathcal H=\half\int_{S^1}\la \partial_\sigma g\,g^{-1},\BE\, \partial_\sigma g\,g^{-1}\ra \,d\sigma$$ 
where $\BE:\g\to\g$ is the map with $+1$-eigenspace $\CE_+$ and $-1$-eigenspace $\CE_-$. 
One can then write the action functional in the ``$\int p\,dq -\mathcal H\,dt$''-form \cite{KS2}
\begin{equation}\label{pdq-Hdt}
S(g)=\half\int_{\Sigma}\la\partial_\sigma g\,g^{-1},\partial_\tau g\,g^{-1}\ra\,d\sigma\,d\tau+\int_Y g^*\eta  -\half\int_\Sigma\la\partial_\sigma g\,g^{-1},\BE\, \partial_\sigma g\,g^{-1}\ra \,d\sigma\,d\tau
\end{equation}
$$ 
\begin{tikzpicture}
\node at(-0.8,1) {$Y$};
\draw[thick] (0,1) ellipse [x radius=0.3, y radius=1];
\draw[thick] (0,0)--(4,0) (0,2)--(4,2)
(4,0) arc [start angle=-90, end angle=90,
x radius=0.3, y radius=1]; 
\draw[dashed] (4,2) arc [start angle=90, end angle=270,
x radius=0.3, y radius=1]; 
\draw[->-=0.8] (0.24,0.4)--node[near start,above]{$\tau$}(4.24,0.4);
\draw[->-=0.6] (2,2) arc [start angle=90, end angle=-90,
x radius=0.3, y radius=1] (2.5,1.6)node{$\sigma$};
\end{tikzpicture}
$$
where $Y$ a solid cylinder with $\Sigma$ as the boundary tube, and $\eta\in\Omega^3(G)$ is the bi-invariant closed 3-form given by 
\begin{equation}\label{eta}
\eta(u,v,w)=\half\la u,[v,w]\ra\quad(\forall u,v,w\in\g).
\end{equation}
 This gives a duality-invariant description of the problem. The Hamiltonian point of view was further developed in \cite{CM,Stern}.
\vskip 1em

In this work we present another duality-invariant description, which is ``holographic'' in the spirit: the $\sigma$-models are equivalent to the Chern-Simons theory on the solid cylinder $Y$, with the (non-topological) boundary condition 
$$*(A|_\Sigma)=\BE(A|_\Sigma).$$
  Besides being explicitly Lorentz-invariant (unlike the Hamiltonian description), this description opens new possibilities for  development of Poisson-Lie T-duality. In particular, it can be seen as a continuous version of the (largely conjectural) quantum Kramers-Wannier duality from \cite{KW}, which takes place on the boundary of the Reshetikhin-Turaev-Witten TQFT corresponding to a suitable quasi-triangular Hopf algebra.

The idea of a 2-dimensional $\sigma$-model appearing on the boundary of   Chern-Simons theory is certainly not new: the best-known example is Witten's observation \cite{Witten} that the WZW-model appears in this way. Our treatment is very similar to \cite{MooreSeiberg}, the main difference comes from different boundary conditions.

Let us briefly discuss what has to be changed on the above picture in the presence of spectator coordinates. In the Abelian case T-duality is an equivalence of the $\sigma$-models given by two torus fibrations over the same base. In the Poisson-Lie case one needs a principal $G$-bundle $P\to P/G$ with vanishing first Pontryagin class. If $H,H'\subset G$ are Lagrangian subgroups then Poisson-Lie T-duality gives an equivalence of the target spaces $P/H$ and $P/H'$. Chern-Simons theory needs to be replaced by the so called Courant 
$\sigma$-model using a certain transitive Courant algebroid over $P/G$. 

Exact Courant algebroids and their reduction were introduced in \cite{PLC} as the geometric structure behind (Poisson-Lie) T-duality. These ideas were rediscovered and extended in \cite{red} and \cite{CG}. The basic idea of the present paper, namely that 2-dimensional $\sigma$-models emerge on the boundary of 3-dimensional Courant $\sigma$-models, makes this link much less mysterious.

These examples suggest a natural generalization: to consider the AKSZ models \cite{AKSZ} (Chern-Simons and Courant are AKSZ models in dimension 3) with appropriate non-topological boundary conditions and to see what kind of (non-topological) models arise on the boundary and which dualities we can see in this way. This generalization will be treated in a future work.

\acknowledgments{Supported in part by  the grant MODFLAT of the European Research Council and the NCCR SwissMAP of the Swiss National Science Foundation.}

\section{Classical boundary conditions of  Chern-Simons theory}\label{sec:cs-bdr}

\subsection{Chern-Simons action}
Let us recall some basic properties of the Chern-Simons action functional.

Let $\g$ be a Lie algebra with an invariant inner product $\la,\ra$, $G$ a connected Lie group integrating $\g$, and $Y$ a compact oriented  3-manifold, possibly with boundary. If $A\in\Omega^1(Y,\g)$ is a $\g$-connection on $Y$, its Chern-Simons action is
\begin{equation}\label{SCS}
S(A)=\int_Y\frac{1}{2}\la A,dA\ra +\frac{1}{6}\la A,[A,A]\ra.
\end{equation}

The variation of $S$ is 
$$\delta S(A)=\int_Y\la\delta A,F_A\ra+\half\int_{\partial Y}\la\delta A,A\ra$$
where $F_A=dA+[A,A]/2$ is the curvature of $A$. 
The boundary term
$$\theta_{\partial Y}:=\half\int_{\partial Y}\la\delta A,A\ra$$
of $\delta S$ is a 1-form $\theta_{\partial Y}$ on the space $\Omega^1(\partial Y,\g)$, and $\omega_{\partial Y}:=\delta\theta_{\partial Y}$ is the Atiyah-Bott symplectic form on $\Omega^1(\partial Y,\g)$ (as is usual, we use $\delta$ as the notation of the de Rham differential on an infinite-dimensional space)
$$\omega_{\partial Y}=-\half\int_{\partial Y}\la\delta A,\delta A\ra,\quad\omega_{\partial Y}(\alpha,\beta)=-\int_{\partial Y}\la\alpha,\beta\ra.$$

Under a gauge transformation $A^g=g^{-1}dg+g^{-1}Ag$ ($g:Y\to G$) the action transforms as
\begin{equation}\label{gauge-trafo}
S(A^g)=S(A)-\int_Y g^*\eta+\int_{\partial Y}\la A,dg\,g^{-1}\ra
\end{equation}
where $\eta\in\Omega^3(G)$ is the bi-invariant closed 3-form given by \eqref{eta}. As a result, $\exp(iS(A))$ is  invariant under gauge transformations trivial on $\partial Y$ provided the periods of $\eta$ are multiples of $2\pi$. The amplitude $\exp(iS(A))$ then makes sense for a connection $A$ on a principal $G$-bundle $P\to Y$, provided a trivialization of $P$ over $\partial Y$ is chosen, and provided there exists an extension of this trivialization to the entire $P\to Y$. The latter condition is always satisfied for 1-connected $G$'s; for other $G$'s extra care is needed (see \cite{DW}), and we will ignore possible resulting problems in this paper.

\subsection{The main boundary condition}\label{sec:Econd}

Let us now consider a boundary condition which makes the boundary term $\theta_{\partial Y}(\delta A)=\frac{1}{2}\int_{\partial Y}\la\delta A,A\ra$ of $\delta S$ vanish, and which, as we shall see later, makes  Chern-Simons theory equivalent to a 2-dimensional $\sigma$-model. 

Let 
$$\BE:\g\to\g$$
be a reflection, i.e.\ a  linear map such that
$$\BE^2=1,\quad\la \BE u,\BE v\ra=\la u,v\ra,$$
with the additional properties 
$$\on{Tr}\BE=0,\quad \la u,\BE u\ra >0\ (\forall\, u\in\g,\;u\neq0).$$
 The map $\BE$ is called (in the context of generalized complex geometry) a  generalized metric \cite{Marco}. It is equivalent to a choice of a half-dimensional subspace $\CE_+\subset\g$ such that $\la,\ra|_{\CE_+}$ is positive-definite and $\la,\ra|_{\CE_+^\perp}$ is negative-definite: $\CE_+$ is the $+1$-eigenspace of $\BE$ and $\CE_-:=(\CE_+)^\perp$ its $-1$-eigenspace.

 Let us choose a pseudo-conformal structure on $\partial Y$ and impose the boundary condition
\begin{equation}\label{bcond}
*(A|_{\partial Y})=\BE(A|_{\partial Y})
\end{equation}
where $*:\Omega^1(\partial Y)\to\Omega^1(\partial Y)$ is the Hodge star. In local isotropic coordinates $t_+,t_-$ on $\partial Y$ we have $*dt_+=dt_+$, $*dt_-=-dt_-$; if 
$$A|_{\partial Y}=a_+dt_+ + a_-dt_-,$$ the boundary condition says 
$$a_+\in \CE_+,\ a_-\in \CE_-.$$

The boundary condition \eqref{bcond} implies that the boundary term $\half\int_{\partial Y}\la\delta A,A\ra$ of $\delta S$ vanishes, as it makes  $\la\delta A|_{\partial Y},A|_{\partial Y}\ra\in\Omega^2(\partial Y)$ vanish.
As a result, solutions of the Euler-Lagrange equations are flat connections on $Y$ satisfying the boundary condition. Notice that if $A=-dg\,g^{-1}$ for a map $g:Y\to G$ then $g|_{\Sigma}$ satisfies \eqref{eq-motion}.

Let us note that for a generic $\BE$ our system is invariant only under the gauge transformations vanishing at $\partial Y$. More precisely we should thus say that solutions of equations of motions are flat connections on $Y$ satisfying the boundary condition, modulo gauge transformations vanishing on $\partial Y$.

\subsection{A topological boundary condition}\label{sec:topo-bc}
Suppose that $\h\subset\g$ is a Lagrangian Lie subalgebra, i.e.\ that $\h^\perp=\h$, and let us impose the condition
$$A|_{\partial Y}\in\Omega^1(\partial Y,\h)\subset\Omega^1(\partial Y,\g).$$
This condition again makes the boundary term of $\delta S$ vanish.

Let $H\subset G$ be the connected subgroup integrating $\h\subset\g$; let us suppose that $H$ is closed. Let us consider gauge transformations 
$$g:Y\to G\text{ such that }g(\partial Y)\subset H.$$
 Equation \eqref{gauge-trafo} then shows that $\exp(iS(A))$ is invariant under these transformations provided the relative cohomology class $[\eta]\in H^3(G,H;\R)$ lies in the image of $H^3(G,H;2\pi\mathbb Z)$ (the class $[\eta]$ is well defined, as $\eta|_H=0$).

A more invariant version of this boundary condition is as follows: we have a principal $G$-bundle $P\to Y$, a reduction of $P$ over $\partial Y$ to a principal $H$-bundle (i.e.\ a submanifold $Q\subset P$ which is a principal $H$-bundle $Q\to\partial Y$), and consider only connections on $P$ which are compatible with the reduction (i.e.\ which restrict to a connection on $Q$). 
 With this boundary condition Chern-Simons theory remains (at least on the classical level) a topological theory.

\subsection{General boundary conditions}
Let us now discuss more general boundary conditions (b.c.'s) given by exact Lagrangian submanifolds of $\Omega^1(\partial Y,\g)$. We shall consider only those b.c.'s that don't depend on  derivatives of $A$. This section is not needed for the rest of the paper, but it is useful for understanding of the general structure. We leave as an exercise to the interested reader to extend the remainder of this paper to these more general boundary conditions.

 Let $T_x:=T_x(\partial Y)$ be the tangent space at $x\in\partial Y$. On the vector space $W_x:=T_x^*\otimes\g$ we have a (constant) symplectic form
$$\omega_x:\twedge^2 W_x\to\twedge^2 T_x^*$$
with values in the 1-dimensional vector space $\twedge^2 T_x^*$, given by
$$\omega_x(\alpha\otimes u,\beta\otimes v)=-\la u,v\ra\,\alpha\wedge\beta,$$
and also a (non-constant) 1-form $\theta_x$ such that $d\theta_x=\omega_x$, given by
$$\theta_x=\half i_{\varepsilon_x}\omega_x$$
where $\varepsilon_x$ is the Euler vector field on the vector space $W_x$. 
They are ``pointwise versions'' of $\omega_{\partial Y}$ and $\theta_{\partial Y}$.

Let now $L_x\subset W_x$ be an exact Lagrangian submanifold, i.e.\ $L_x$ is Lagrangian and  there is a $\twedge^2T^*_x$-valued function $f_x$ on $L_x$ such that $\theta_x|_{L_x}=df_x$. Let us also suppose that $L_x$'s depend smoothly on $x$ in the sense that their union is a smooth submanifold $L$ of $W:=T^*(\partial Y)\otimes\g$, and also that $f_x$ depends smoothly on $x$, i.e.\ that $f_x$'s combine to a smooth map $f:L\to\twedge^2T^*\partial Y$.

The b.c.\ we impose on $A|_{\partial Y}$ is that it is a section of $L\mapsto \partial Y$ (the b.c.'s considered above are of this form, with $f=0$). We need to add a boundary term
$$S_\partial(A):=\int_{\partial Y}f(A|_{\partial Y})$$
to the action $S(A)$. The variation of the action $S+S_\partial$ (under the b.c.) is
$$\delta(S+S_\partial)=\half\int_Y\la\delta A,F_A\ra,$$
i.e.\ the boundary term of the variation disappears.

\section{From Chern-Simons to a Hamiltonian system}\label{sec:CSHam}
\subsection{Chern-Simons action on a cylinder}\label{Csfcyl}
Let $\disk$  be a disk, $I$ an interval, and $Y=\disk\times I$. Let $\Sigma=(\partial\, \disk)\times I\subset \partial Y$. On $\Sigma$ we impose the boundary condition \eqref{bcond}. We use the standard pseudo-Riemannian metric $-d\tau^2+d\sigma^2$ on $\Sigma$,   
 where $\tau$ is the coordinate on $I$ and $\sigma$ is the angle along the circle $S^1=\partial \,\disk$. In these coordinates we have 
$$*d\tau=d\sigma,\ *d\sigma=d\tau.$$
The boundary condition thus requires $A|_\Sigma$ to be of the form
\begin{equation}\label{bcond-cyl}
A|_\Sigma=a_\sigma\,d\sigma + \BE(a_\sigma)\,d\tau,\quad a_\sigma:\Sigma\to\g.
\end{equation}

Let us now analyze the Chern-Simons action \eqref{SCS} on $Y=\disk\times I$ with the boundary condition \eqref{bcond-cyl}. Let us use the notation
$$A=a_\tau\,d\tau+\tilde A\quad (a_\tau:Y\to\g),$$
where $\tilde A$ is a $\tau$-dependent $\g$-valued 1-form on $\disk$, and 
$$F_{\tilde A}:=d_h\tilde A+[\tilde A,\tilde A]/2$$
where $d_h$ is the de Rham differential on $\disk$ (as opposed to $Y$). A simple calculation gives
$$\frac{1}{2}\la A,dA\ra +\frac{1}{6}\la A,[A,A]\ra=d\tau\wedge\Bigl(\la a_\tau,F_{\tilde A}\ra+\half\bigl\la\frac{\partial\tilde A}{\partial \tau},\tilde A\bigr\ra\Bigr)+\half d\la a_\tau\,d\tau,\tilde A\ra$$
and thus
\begin{equation}\label{S-disk}
S(A)=\int_I\,\biggl(\,\int_{\sdisk} \la a_\tau,F_{\tilde A}\ra+\half\int_{\sdisk}\bigl\la\frac{\partial\tilde A}{\partial \tau},\tilde A\bigr\ra-\half\int_{\partial \,\sdisk}\bigl\la a_\sigma,\BE(a_\sigma)\bigr\ra d\sigma\,\biggr)\,d\tau.
\end{equation}

The first term in \eqref{S-disk} (with $a_\tau$ a Lagrange multiplier) imposes the constraint $F_{\tilde A}=0$. If we set $\tilde A=-d_hg\,g^{-1}$, where $g$ is a map $g:\disk\times I\to G$, then the action \eqref{S-disk} gets equal to the action $S(g)$ given by \eqref{pdq-Hdt}. \emph{We thus recovered the Hamiltonian description of  Poisson-Lie T-duality.}

\subsection{Chern-Simons on a cylinder as a Hamiltonian system}
Let us now discuss the meaning of 
 the action \eqref{S-disk} in more detail. As we observed, in the first term $a_\tau$ is a Lagrange multiplier, i.e. it gives us a constraint 
$$F_{\tilde A}=0.$$
The remaining terms depend only on $\tilde A$, which is a (time-dependent) flat $\g$-connection on $\disk$. Taking into account the gauge invariance of the action \eqref{S-disk} under gauge transformations vanishing on the boundary, we should rather consider gauge classes of flat connections $\tilde A$, i.e.\ their moduli space $\mathcal M_G(\disk,\partial\,\disk)$. The space $\mathcal M_G(\disk,\partial\,\disk)$ is the subspace of $\Omega^1(S^1,\g)$ given by the unit holonomy constraint.

The second term in \eqref{S-disk} is
\begin{equation}\label{S-2nd}
\int_I (\tilde A)^*\theta_{\,\sdisk}
\end{equation}
where we understand $\tilde A$ as a map $I\to\Omega^1(\disk,\g)$ and $\theta_{\,\sdisk}$ is the 1-form on $\Omega^1(\disk,\g)$ given by
$$\theta_{\,\sdisk}=\half\int_\sdisk\la\delta A,A\ra.$$
 The symplectic form $\omega_{\,\sdisk}=\delta\theta_{\,\sdisk}$ on $\Omega^1(\disk,\g)$, when restricted to the subspace of flat connections, becomes degenerate, but it descends to a symplectic form (of Atiyah and Bott) on $\mathcal M_G(\disk,\partial\,\disk)$. The term \eqref{S-2nd} thus means that we have a Hamiltonian system on the symplectic manifold $\mathcal M_G(\disk,\partial\,\disk)$, with the Hamiltonian given by the third term of \eqref{S-disk},
\begin{equation}\label{Ham}
\mathcal H=\half\int_{\partial\,\sdisk}\bigl\la a_\sigma,\BE(a_\sigma)\bigr\ra d\sigma.
\end{equation}

\subsection{Chern-Simons on a hollow cylinder as a Hamiltonian system}\label{sec:CShollow}

Let $\annu$  be an annulus, and let let us consider Chern-Simons action for $Y^*=\annu\times I$. On the outer cylinder $\Sigma_\text{out}=S^1\times I$ we impose the same boundary condition $*(A|_{\Sigma_\text{out}})=\BE(A|_{\Sigma_\text{out}})$ as before, while on the inner cylinder $\Sigma_
\text{inn}$ we impose the condition 
$$A|_{\Sigma_\text{inn}}\in\Omega^1(\Sigma_\text{inn},\h),$$
or more generally (as discussed in Section \ref{sec:topo-bc}), we choose a reduction of $P$ over $\Sigma_\text{inn}$ to a principal $H$-bundle, and consider only compatible connections on $P$.

We then get
$$S(A)=\int_I\,\biggl(\,\int_{\sannu} \la a_\tau,F_{\tilde A}\ra+\half\int_{\sannu}\bigl\la\frac{\partial\tilde A}{\partial \tau},\tilde A\bigr\ra-\half\int_{S^1}\bigl\la a_\sigma,\BE(a_\sigma)\bigr\ra\, d\sigma\,\biggr)\,d\tau.$$
We thus still have a Hamiltonian system, but with a slightly larger phase space $\mathcal M_{G,H}$ defined as follows: 
$\mathcal M_{G,H}$ is the moduli space of flat $\g$-connection on the annulus $\annu$, which restrict to a $\h$-valued 1-form on the inner boundary circle, modulo gauge transformations vanishing on the outer circle and taking values in $H\subset G$ on the inner circle.

As we shall see, $\mathcal M_{G,H}$ can be interpreted as the cotangent bundle of the free loop space $L(G/H)$ (twisted by a closed 3-form on $G/H$), and the Hamiltonian system is equivalent to a $\sigma$-model with the target $G/H$.

Notice that $\mathcal M_G(\disk,\partial\,\disk)$ is obtained from $\mathcal M_{G,H}$ by symplectic reduction: if we restrict the holonomy along the inner circle to be trivial, which gives a coisotropic submanifold in $\mathcal M_{G,H}$, and we mod out by its null leaves, we obtain $\mathcal M_G(\disk,\partial\,\disk)$.

The Hamiltonian system on $\mathcal M_G(\disk,\partial\,\disk)$ (Chern-Simons on the full cylinder), is duality-invariant (i.e.\ $H$-independent) part of the Hamiltonian system on $\mathcal M_{G,H}$, corresponding to the non-Abelian momentum constraint.

\section{The Hamiltonian system as a $\sigma$-model}\label{sec:ham-is-sigma}

In this section we shall show that the Hamiltonian system on $\mathcal M_{G,H}$ with the Hamiltonian $\mathcal H$ (i.e.\ Chern-Simons theory on a hollow cylinder with the boundary conditions described in Section \ref{sec:CShollow}) is equivalent to a 2-dimensional $\sigma$-model with the target space $G/H$.  A  conceptual explanation uses Courant $\sigma$-models and reduction of Courant algebroids; we postpone it to Section \ref{sec:courant}, where we also deal with spectator coordinates.

\subsection{$G/H$ as the target of a 2-dimensional $\sigma$-model}\label{sec:GH-sigma}

Let us choose an auxiliary connection $\mathcal A\in\Omega^1(G,\h)$ on the principal $H$-bundle $G\to G/H$. The $H$-invariant closed 3-form
\begin{equation}\label{etaA}
\eta_{\mathcal A}:=\eta+\half\,d\la\mathcal A,g^{-1}dg\ra
\end{equation}
is basic, i.e.\ it is the pullback of a closed 3-form from 
$G/H$, which we will also denote by $\eta_{\mathcal A}$. 

Let us suppose that the horizontal spaces of the connection $\mathcal A$ are Lagrangian w.r.t.\ the inner product $\la,\ra$; we shall say that such a connection is \emph{Lagrangian}. (There is a canonical choice for a Lagrangian connection: we extend the inner product $(u,v)_\BE:=\la u,\BE v\ra$ on $\g$ to a right-invariant Riemann metric on $G$, and let $\mathcal A$ be the connection whose horizontal spaces are $(,)_\BE$-perpendicular to the vertical spaces.)

The Lagrangian connection $\mathcal A$ can now be used to identify the bundle $TG$ with $p^*\bigl((T\oplus T^*)(G/H)\bigr)$, where $p:G\to G/H$ is the projection: the horizontal subbundle of $TG$ is identified with $T(G/H)$ and the vertical with $T^*(G/H)$ (via $\la,\ra$). The subbundle $(\CE_+)^R\subset TG$ (the right-translated $\CE_+$) thus becomes the graph of a bilinear form
$$E_{\mathcal A}\in\Gamma\bigl((T^*)^{\otimes 2} (G/H)\bigr).$$

The symmetric part of the tensor field $E_{\mathcal A}$ is a Riemann metric. If $\mathcal A,\mathcal A'$ are two Lagrangian connections then, as an easy calculation shows,
$$E_{\mathcal A'}=E_{\mathcal A}-B,\quad\eta_{\mathcal A'}=\eta_{\mathcal A}+dB$$
for some 2-form $B$. ($E_{\mathcal A}$ is symmetric  iff  $\mathcal A$ is the canonical Lagrangian connection.)

We can use $E_{\mathcal A}$ and $\eta_{\mathcal A}$ to define a $\sigma$-model with the target $G/H$, with the standard action functional (for $f:Y\to G/H$, where $Y$ is a solid cylinder with the boundary $\Sigma$)
\begin{equation}\label{sigmaGH}
    S(f)=\int_\Sigma E_{\mathcal A}(\partial_+f,\partial_-f) +\int_Y f^*\eta_{\mathcal A}
\end{equation}
where $df=\partial_+f+\partial_-f$ is the splitting of $df$ to the $(1,0)$ and the $(0,1)$-components.

\subsection{The Hamiltonian system is the $\sigma$-model with target $G/H$}

Let us now explain why the Hamiltonian system on $\mathcal M_{G,H}$ is, in fact, the $\sigma$-model with the target space $G/H$, given by the tensor field $E_{\mathcal A}$ and by the closed 3-form $\eta_{\mathcal A}$. First of all, we need to identify the phase space $\mathcal M_{G,H}$ with $T^*L(G/H)$.

The moduli space $\mathcal M_{G,H}$ can be naturally identified with the space of quasi-periodic maps
$$g:\R\to G\quad\text{such that}\quad g(x+2\pi)=g(x)h\ \text{for some}\ h\in H$$
modulo right multiplication by elements of $H$. Indeed, for any $[A]\in\mathcal M_{G,H}$ (where $A\in\Omega^1(\annu,\g)$ is a flat connection) we choose a map $g:\widetilde{\annu}\to G$ such that $A=-dg\,g^{-1}$ (where $\widetilde{\annu}$ is the universal cover of the annulus $\annu$) and such that $g$ takes values in $H$ on (the cover of) the inner circle of $\annu$; restricting $g$ to the cover of the outer circle we get a quasi-periodic map as needed. 

We can now identify $\mathcal M_{G,H}$ with $T^*(L(G/H))$: $g$ project to a map $f:S^1\to G/H$, i.e.\ to an element of $L(G/H)$, and the vertical part (w.r.t.\ $\mathcal A$) of $dg$ is a $f^*(T^*(G/H))$-valued 1-form on $S^1$, i.e.\ gives us a covector at $f\in L(G/H)$.

A straightforward calculation now shows that the identification 
$$\mathcal M_{G,H}\cong T^*(L(G/H))$$
 is a symplectomorphism provided we add to the standard symplectic form on $T^*(L(G/H))$ the closed 2-form on $L(G/H)$ obtained from $-\eta_{\mathcal A}$ by integration over $S^1$ (i.e.\ we twist the cotangent bundle) and that the Hamiltonian $\mathcal H$ coincides with the Hamiltonian of the $\sigma$-model \eqref{sigmaGH}. In place of doing these calculations here we present a  conceptual explanation in Section \ref{sec:exsigGH}. 

\section{Courant algebroids and dg symplectic manifolds}
In this section we summarize some basic definitions and facts concerning Courant algebroids.

\subsection{Exact Courant algebroids}
Courant algebroids, introduced in \cite{lxw}, are a generalization of Lie algebras with invariant inner product. By definition, a \emph{Courant algebroid}  (CA) is a vector bundle $V\to M$ endowed with an inner product $\la\,,\,\ra_V$ on its fibres, with a vector bundle map $a_V:V\to TM$ called the \emph{anchor}, and with a bilinear map $[\,,\,]_V:\Gamma(V)\times\Gamma(V)\to\Gamma(V)$ such that for all $u,v,w\in\Gamma(V)$ and $f\in C^\infty(M)$
\begin{align*}
[u,[v,w]_V]_V&=[[u,v]_V,w]_V+[v,[u,w]_V]_V\\
a_V\bigl([u,v]_V\bigr)&=[a_V(u),a_V(v)]\\
[u,fv]_V&=f[u,v]_V+(a_V(u)f)v\\
a(u)\la v,w\ra_V&=\la[u,v]_V,w\ra_V+\la v,[u,w]_V\ra_V\\
[u,u]_V&=a_V^t\bigl(d\la u,u\ra_V/2\bigr)
\end{align*}
where
$a_V^t:T^*M\to E^*\xrightarrow{\la,\ra_V} E$
is the transpose of $a_V$.

A CA is \emph{exact} if the sequence
\begin{equation}\label{exact}
0\to T^*M\xrightarrow{a_V^t}V\xrightarrow{a_V}TM\to0
\end{equation}
is exact. As shown in \cite{PLC}, exact CAs are classified by $H^3(M,\R)$: If we split the exact sequence \eqref{exact} so that $TM\subset TM\oplus T^*M\cong V$ is $\la\,,\,\ra_V$-isotropic then the 3-form $\eta\in\Omega^3(M)$ given by
$$\eta(u,v,w):=\la[u,v]_V,w\ra_V\quad\forall u,v,w\in\Gamma(TM)\subset\Gamma(V)$$
is closed and its cohomology class is independent of the splitting. The Courant bracket $[,]_V$ on $V\cong TM\oplus T^*M$ is
$$[(u,\alpha),(v,\beta)]_V=\bigl([u,v],L_u\beta-i_v\alpha+\eta(u,v,\cdot)\bigr)\quad\forall u,v\in\Gamma(TM),\alpha,\beta\in\Gamma(T^*M).$$

\subsection{CAs and dg symplectic manifolds}
Courant algebroids are equivalent to non-negatively graded manifolds with a symplectic form of degree $2$ and with a function $\Theta$ of degree 3 satisfying the classical master equation $\{\Theta,\Theta\}=0$ \cite{royth,some}. Namely, if $\mathcal V$ is such a graded manifold then the vector bundle $V\to M$ is given by $\Gamma(V)=C^\infty(\mathcal V)^1$, $C^\infty(M)=C^\infty(\mathcal V)^0$, with the Courant algebroid structure 
\begin{align*}
[u,v]_V&=\{\{\Theta,u\},v\}\\
\la u,v\ra_V&=\{u,v\}\\
a_V(u)f&=\{\{\Theta,u\},f\}
\end{align*}
In local coordinates   $x^i,\deg x^i=0$, $e^a,\deg e^a=1$, $p_i,\deg p_i=2$, such that the symplectic form on $\mathcal V$ is 
\begin{equation}\label{darb}
\omega=dp_i\,dx^i+g_{ab}\,de^a\,de^b
\end{equation}
for some constant symmetric matrix $g_{ab}$, we have
\begin{equation}\label{Theta}
\Theta=a_a^i(x)p_ie^a-\frac{1}{6}c(x)_{abc}e^ae^be^c
\end{equation}
where $a_V(e_a)=a^i_a\partial/\partial x^i$ and $c_{abc}=\la[e_a,e_b]_V,e_c\ra$ (here $e_a=g_{ab}e^b$).

In particular, if $V\to M$ is an exact Courant algebroid, the corresponding $\mathcal V$ is $T^*[2]T[1]M$, with $\Theta=d+\eta$ (where the de Rham differential $d$, which is a vector field on $T[1]M$, is seen as a function on $T^*[2]T[1]M$). In local coordinates $x^i$, $\xi^i=dx^i$, $\pi_i$, $p_i$, we have $\omega=dp_i\,dx^i+d\pi_i\,d\xi^i$ and $\Theta=p_i\xi^i-\frac{1}{6}\eta_{ijk}(x)\xi^i\xi^j\xi^k$. 

\subsection{Equivariant CAs and reduction}\label{sec:equivCA}

If $\g$ is a Lie algebra with a (possibly degenerate) invariant symmetric bilinear pairing $\la,\ra_\g$, a \emph{$(\g,\la,\ra_\g)$-equivariant CA} is a CA $V\to M$ together with a linear map $\rho:\g\to\Gamma(V)$ satisfying 
$$[\rho(u),\rho(v)]_V=\rho([u,v]),\quad\la\rho(u),\rho(v)\ra_V=\la u,v\ra_\g.$$
The derivations $[\rho(u),\cdot]_V$ give in this case an action of $\g$ on $V$. If this action integrates to an action of a connected group $G$ with the Lie algebra $\g$, we shall say that $V$ is $(G,\la,\ra_\g)$-equivariant.

Equivariant exact CAs can be classified in the case of free and proper actions \cite{PLC}. Suppose that $G$, a connected Lie group with Lie algebra $\g$, acts freely and properly on $P$, i.e.\ that $P\to P/G$ is a principal $G$-bundle. Let us choose a connection $A$ on this principal $G$-bundle and let $F_A$ be its curvature. Then there is a bijection between isomorphism classes of $G$-equivariant exact CAs $V\to P$ and solutions of
\begin{equation}\label{pontr}
d\alpha=\half\la F_A,F_A\ra_\g,\quad\alpha\in\Omega^3(M/G)
\end{equation}
modulo exact 3-forms on $P/G$. In particular, $V\to P$ exists iff the Pontryagin class $[\la F_A,F_A\ra]\in H^4(P/G,\R)$ vanishes. Explicitly, if $\alpha$ is a solution of \eqref{pontr} then $V=(T\oplus T^*)P$ with the closed 3-form
$$\eta=\alpha-\Bigl(\frac{1}{2}\la A,dA\ra_\g +\frac{1}{6}\la A,[A,A]\ra_\g\Bigr)$$
and with 
$$\rho(u)=\bigl(u_P,\half\la u,A\ra_\g\bigr)$$
is the corresponding equivariant exact CA over $P$.

Equivariant CAs can be reduced in the following way \cite{red,PLC}. If  $V\to P$ is a $G$-equivariant CA as above, let  
$$(V_{/G})_x:=(\rho_x(\g))^\perp/\rho_x(\g')\quad(\forall x\in P)$$
where $\g'\subset\g$ is the kernel of $\la,\ra_\g$. After taking quotient by $G$, $V_{/G}$ becomes a vector bundle $V_{/G}\to P/G$, and the CA structure on $V\to P$ descends to a CA structure on $V_{/G}\to P/G$. If $V$ is exact and $\la,\ra_\g=0$ then $V_{/G}$ is also exact; for general $\la,\ra_\g$ the CA $V_{/G}$ is only transitive (i.e.\ its anchor map is surjective).

The reduction of CAs can be seen as a symplectic reduction \cite{some}. The cone $C\g$ of $\g$ ($C\g$ is the differential graded Lie algebra $C\g=\g[\epsilon]$ with $\deg\epsilon=-1$ (and thus $\epsilon^2=0$) and $d\epsilon=1$) has a central extension
$$0\to\R[2]\to\tilde C\g\to C\g\to0$$
 given by  
$$[u\epsilon,v\epsilon]=\la u,v\ra_\g\,s$$
where $s=1\in\R[2]$ is the generator of the center, and $ds=0$. A CA $V\to P$ is $\g$-equivariant iff the corresponding dg symplectic manifold $\mathcal V$ is equipped with a dg Poisson (i.e.\ moment) map
$$\mu:\mathcal V\to (\tilde C\g)^*[2]$$
such that $\la\mu,s\ra=1$. The reduction of $V$ is then equivalent to the symplectic reduction of $\mathcal V$ at the moment value $(1,0)\in \R[2]\oplus(C\g)^*[2]=(\tilde C\g)^*[2]$.

\section{Spectator coordinates and Courant $\sigma$-models}\label{sec:courant}
In this section we shall see how replacing Chern-Simons with more general Courant $\sigma$-models we can get Poisson-Lie T-duality with spectator coordinates, and how it gives a more conceptual explanation of what we did above. A central role is played by a link between exact Courant algebroids and 2-dimensional $\sigma$-models.

\subsection{AKSZ  and Courant $\sigma$-models}
Let us briefly recall the AKSZ models introduced in \cite{AKSZ}. Suppose that $\mathcal V$ is a graded manifold with a symplectic form $\omega$ of degree $n>0$ and that $\Theta$ is a function on $\mathcal V$ of degree $n+1$ such that $\{\Theta,\Theta\}=0$; in particular $Q:=\{\Theta,\cdot\}$ is a homological vector field on $\mathcal V$. Let 
$$\theta:=\frac{1}{n}i_\varepsilon\omega$$
where $\varepsilon$ is the Euler vector field given by $\varepsilon f=(\deg f) f$ for any homogeneous function $f$ on $\mathcal V$. Since $L_\varepsilon\,\omega=n\,\omega$, we have $d\theta=\omega$.

For any oriented compact $n+1$-manifold $Y$ and any map $F:T[1]Y\to \mathcal V$ the AKSZ action functional is
$$S(F)=\int_{T[1]Y}i_{d}(F^*\theta)-F^*\Theta$$
(where $d$ is the de Rham differential on $Y$, seen as a vector field on $T[1]Y$; the integral over $T[1]Y$ is the usual integral of differential forms over $Y$). The stationary points of $S$ are the differential graded maps $T[1]Y\to \mathcal V$. 
(While the main idea of \cite{AKSZ} is BV quantization of this theory, we shall simply consider the action functional $S$ for grading-preserving maps $F$, just as we did for Chern-Simons theory.)

Let us now restrict to the case of $n=2$, and suppose that $\mathcal V$ is non-negatively graded. In this case $\mathcal V$ is equivalent to a Courant algebroid $V\to M$; the corresponding AKSZ model is called the ($V$-)Courant $\sigma$-model \cite{I,roy}. In the local Darboux coordinates $x^i,e^a,p_i$ satisfying \eqref{darb} we have
$$\theta=p_i\,dx^i+\half g_{ab}\,e^a\,de^b.$$
The AKSZ action functional is thus (using the expression \eqref{Theta} for $\Theta$) 
\begin{equation}\label{CAsigma}
S=\int_Y p_i\,dx^i+\half g_{ab}\,e^a\,de^b - a_a^i(x)p_ie^a+\frac{1}{6}c(x)_{abc}e^ae^be^c
\end{equation}
where the fields are  $x^i\in \Omega^0(Y)$, $e^a\in\Omega^1(Y)$, $p_i\in\Omega^2(Y)$ ($\dim Y=3$).

 Chern-Simons theory is a special case of the Courant $\sigma$-model, namely when $V=\g$ (or equivalently, when there are no $x^i$'s and $p_i$'s, only $e^a$'s).

\subsection{Boundary conditions for the Courant $\sigma$-models}\label{sec:bCourM}
The variantion of the action functional \eqref{CAsigma} is
\begin{multline*}
\delta S=\int_{\partial Y}\Bigl( p_i\,\delta x^i-g_{ab}\,e^a\,\delta e^b\Bigr)  \\
+ \int_Y\Bigl( \delta p_i\bigl(dx^i-Q\,x^i\bigr)+g_{ab}\,\delta e^a\bigl(de^b-Q\,e^b\bigr)-\delta x^i\bigl(dp_i-Q\,p_i\bigr)\Bigr).
\end{multline*}
The boundary term of $\delta S$ is a 1-form $\theta_{\partial Y}$ on the space of graded maps $T[1]\partial Y\to\mathcal V$, and $\omega_{\partial Y}:=\delta\theta_{\partial Y}$ is a symplectic form on this space.

As in Section \ref{sec:cs-bdr} we  now impose a boundary condition given by a Lagrangian submanifold on which $\theta_{\partial Y}$ vanishes (or, at least, is exact). Namely, if we choose, as in Section \ref{sec:Econd}, a generalized metric on $V$, i.e.\ a linear transformation $\BE:V\to V$ of the vector bundle $V\to M$ satisfying
$$\BE^2=1,\ \la\BE u,\BE v\ra=\la u,v\ra,\ \on{Tr}\BE=0,\ \la u,\BE u\ra>0\ (\forall0\neq u \in\Gamma(V)),$$
then the boundary condition (where $*$ is the Hodge star given by a pseudo-Riemannian metric on $\partial\Sigma$)
\begin{equation}\label{CAbcond}
*e^a=\BE^a_b\,e^b,\ p_i=\half\, e^a\,g_{ab}\,\frac{\partial\BE^b_c}{\partial x^i}\,e^c
\end{equation}
satisfies our requirements. (We wrote the boundary condition using coordinates on $\mathcal V$, but notice that the condition on $p_i$'s is forced by the condition on $e^a$'s and by vanishing of $\theta_{\partial Y}$, which are coordinate-independent.)

As in Section \ref{sec:CSHam} we can reinterpret the Courant $\sigma$-model on a solid cylinder $Y$  with the boundary conditions \eqref{CAbcond} as a Hamiltonian system. Namely, if we split the forms $e^a$ and $p_i$ to their horizontal and vertical parts (for any form $\beta$ on $Y=\disk\times I$ we have $\beta=\beta_\text{hor}+d\tau\,\beta_\text{vert}$ where $\beta_\text{vert}$, $\beta_\text{hor}$ are $\tau$-dependent forms on $\disk$) then $(e^a)_\text{vert}$ and $(p_i)_\text{vert}$ are Lagrange multipliers forcing $(x^i,(e^a)_\text{hor},(p_i)_\text{hor})$ to be a ($t$-dependent) dg map $T[1]\disk\to\mathcal V$. The Hamiltonian (which appears as the boundary term after integration by parts) is
$$\mathcal H=\half\int_{S^1}g_{ab}(e^a)_\sigma (e^b)_\sigma\,d\sigma$$
where $(e^a)_\sigma\,d\sigma=e^a|_{S^1}$, and the phase space is the space of all dg maps $T[1]\disk\to\mathcal V$ modulo homotopy relative to the boundary (where by homotopy we mean a dg map $T[1](\disk\times I)\to\mathcal V$), with the symplectic form $\omega_{\,\sdisk}$.
We shall denote this phase space by $\mathcal M_{\mathcal V}(\disk,\partial\,\disk)$ or $\mathcal M_{V}(\disk,\partial\,\disk)$.

\subsection{Exact Courant algebroids and 2-dimensional $\sigma$-models}\label{sec:ex-sigma}
In the case of exact Courant algebroids the action \eqref{CAsigma} becomes after integration by parts
$$S=\int_Y \Bigl(p_i(dx^i-\xi^i)+\pi_i\,d\xi^i+\frac{1}{6}\eta_{ijk}(x)\xi^i\xi^j\xi^k\Bigr)+\half\int_\Sigma\pi_i\xi^i.$$
In the bulk integral $p_i$ and $\pi_i$ are Lagrange multipliers imposing $\xi^i=dx^i$ and $d\xi^i=0$. We thus have a map $f:Y\to M$ (with components $x^i$) and the bulk integral is $\int_Y f^*\eta$.

The generalized metric $\BE:V\to V$ is (in our case of $V=(T\oplus T^*)M$) equivalent to a linear map $E:TM\to T^*M$ (with positive-definite symmetric part) whose graph is the $+1$-eigenbundle of $\BE$.
The boundary condition \eqref{CAbcond} is
\begin{align*}
(\pi_i)_+&=E_{ij}(\xi^i)_+\\
(\pi_i)_-&=-E_{ji}(\xi^i)_-
\end{align*}
(we don't write the boundary condition for $p_i$, as it is not needed).
 The boundary part of $S$ is therefore
$\int_\Sigma E_{ij}(\xi^i)_+\,(\xi^j)_-.$ Using the constraint $\xi^i=dx^i$ we thus get
\begin{equation}\label{Ssigma}
S=\int_\Sigma E(\partial_+ f,\partial_-f)+\int_Yf^*\eta
\end{equation}
which is the standard $\sigma$-model action with the target $M$ given by the tensor field $E$ and the closed 3-form $\eta$.

The Hamiltonian approach of Section \ref{sec:bCourM} gives in this case the Hamiltonian description of the $\sigma$-model with the target $M$.

(We should stress that there is a global problem that we didn't solve. The action \eqref{Ssigma}, or rather $\exp(iS)$, should be defined (in the appropriate sense) for maps $\Sigma\to M$ and not require an extension of the map to $Y\supset\Sigma$. A proper treatment should use a global version of  AKSZ models, possibly as in \cite{PTVV}.)

\subsection{The $\sigma$-model with the target $G/H$ revisited}\label{sec:exsigGH}

We can now give a conceptual reason why the Chern-Simons  theory on a hollow cylinder, as studied in Section \ref{sec:CShollow}, is equivalent to the $\sigma$-model with the target $G/H$, described in Section \ref{sec:GH-sigma}. The idea is to use an exact Courant algebroid $V_{G/H}\to G/H$ and the corresponding Courant $\sigma$-model, which, as we just saw, is equivalent to a 2-dimensional $\sigma$-model with the target $G/H$.

We start with the exact CA $V_G\to G$ which is $(\g,-\la,\ra_\g)$-equivariant w.r.t.\ the action of $\g$ on $G$ by the left-invariant vector fields. As follows from Section \ref{sec:equivCA}, the CA $V_G$ is uniquely defined by the equivariance property and explicitly we have $V_G=(T\oplus T^*)G$ with the closed 3-form
$$\eta(u,v,w)=\half\la[u,v],w\ra_\g$$
and the action $\rho(u)=\bigl(u^L,-\la u,\cdot\ra_\g^L\bigr)\in\Gamma\bigl((T\oplus T^*)G\bigr)$ ($\forall u\in\g$).

The reduced Courant algebroid $V_{G/H}:=(V_G)_{/H}\to G/H$ is again exact; its splitting to $(T\oplus T^*)(G/H)$ is equivalent to a choice of a Lagrangian connection $\mathcal A$ on the principal $H$-bundle $G\to G/H$, as in Section \ref{sec:GH-sigma}, and the corresponding closed 3-form on $G/H$ is $\eta_{\mathcal A}$ given by \eqref{etaA}.

 We can also describe $V_{G/H}$ as the trivial bundle $V_{G/H}=\g\times (G/H)$; the Courant bracket on constant sections is the Lie bracket on $\g$, the pairing is  $\la,\ra_\g$, and the anchor map is the action of $\g$ on $G/H$. If we have a generalized metric $\BE:\g\to\g$, it gives us a generalized metric on $V_{G/H}=\g\times (G/H)$ (constant in $G/H$). By the result of Section \ref{sec:ex-sigma} the CA $\sigma$-model given by $V_{G/H}$ with the boundary condition given by $\BE$ is equivalent to the $\sigma$-model \eqref{sigmaGH} with the target $G/H$.

Let us start with Chern-Simons for the full cylinder as in Section \ref{Csfcyl}. (In the end, Chern-Simons on the hollow cylinder can be seen as an auxiliary construction: what it really important is to understand the link between the $V_{G/H}$-Courant $\sigma$-model and the $\g$-Chern-Simons theory on a solid cylinder, i.e.\ why the latter is obtained from the former by the non-Abelian momentum constraint.)

The reduced Courant algebroid $(V_G)_{/G}$ is $(\g,\la,\ra_\g)$ and the corresponding dg symplectic manifold is $\g[1]$. This implies that $\g[1]$ can be obtained by symplectic reduction (i.e.\ by taking a coisotropic dg submanifold and modding out by the null leaves of the restriction of the symplectic form) from the dg symplectic manifold $\mathcal V_{G/H}$ corresponding to the exact CA $V_{G/H}\to G/H$ (it can also be seen from the description of $V_{G/H}$ as $\g\times {G/H}$). As a result, the phase space of the Chern-Simons theory $\mathcal M_{\g}(\disk,\partial\,\disk)$ is obtained from the phase space of the Courant $\sigma$-model $\mathcal M_{V_{G/H}}(\disk,\partial\,\disk)$ by symplectic reduction (by imposing the non-Abelian momentum constraint and modding out the null leaves) and the Hamiltonian descends from $\mathcal M_{V_{G/H}}(\disk,\partial\,\disk)$ to $\mathcal M_{\g}(\disk,\partial\,\disk)$. 

When we consider the hollow cylinder $Y^*$, we  get an isomorphism in place of symplectic reduction, provided we choose appropriate boundary conditions. The boundary condition on the inner tube is, in the Chern-Simons case, given in Section \ref{sec:CShollow}. For the $V_{G/H}$-Courant $\sigma$-model we need to choose a suitable Lagrangian dg submanifold $\mathcal L\subset\mathcal V_{G/H}$ and impose the condition that the restriction of $F:T[1]Y^*\to\mathcal V_{G/H}$ to the inner tube $\Sigma_{inn}\subset\partial Y^*$ is a map $T[1]\Sigma_{inn}\to\mathcal L$. The Lagrangian submanifold $\mathcal L\subset\mathcal V_{G/H}$ is the one given by the Dirac structure $\h\times [1]\subset\g\times (G/H)=V_{G/H}$.

\subsection{Poisson-Lie T-duality with spectator coordinates}

Let  $P\to P/G$ be a principal $G$-bundle and let $V_P\to P$ be a $(G,-\la,\ra_\g)$-equivariant exact CA. Recall from Section \ref{sec:equivCA} that $V_P$ exists iff the characteristic class $[\la F_A,F_A\ra_\g]$ of $P$ vanishes. 

Let  $V_{P/G}\to P/G$ be the reduced CA $V_{P/G}:=(V_P)_{/G}$ and for a closed Lagrangian subgroup $H\subset G$ let $V_{P/H}\to P/H$ be the reduced CA $(V_P)_{/H}$; notice that $V_{P/H}$ is exact ($V_{P/G}$ is only transitive, i.e.\ its anchor is surjective).

We choose a generalized metric $\BE$ on $V_{P/G}$. By construction we have a natural isomorphism of vector bundles $V_{P/H}\cong p^*V_{P/G}$ where $p:P/H\to {P/G}$ is the projection; as a result, $\BE$ gives us a generalized metric on $V_{P/H}$. Since $V_{P/H}$ is exact, we get a $\sigma$-model with the target $P/H$ which is equivalent to the $V_{P/H}$-Courant $\sigma$-model on a solid cylinder with the boundary condition given by $\BE$.

Let us choose another Lagrangian subgroup $H'\subset G$. Poisson-Lie T-duality is an equivalence of the $\sigma$-models with the targets $P/H$ and $P/H'$, i.e.\ of the Courant $\sigma$-models with the CAs $V_{G/H}$ and $V_{G/H'}$, and with the boundary condition given by $\BE$.

The equivalence is given simply by the $V_{P/G}$-Courant $\sigma$-model with the boundary condition given by $\BE$. Indeed, the phase space $\mathcal M_{V_{P/G}}(\disk,\partial\,\disk)$ of the $V_{P/G}$-Courant $\sigma$-model is a reduction of both $\mathcal M_{V_{P/H}}(\disk,\partial\,\disk)$ and of $\mathcal M_{V_{P/H'}}(\disk,\partial\,\disk)$, and the Hamiltonians match. Notice that we cannot say that we have an isomorphism of Hamiltonian systems; we get an isomorphism only when after the reduction, i.e.\ only after we impose the non-Abelian momentum constraint. 
(Let us remark that one can also introduce an inner tube and impose a boundary condition given by a Dirac structure, as in Section \ref{sec:CShollow}, to get a closer link between the $V_{P/G}$-Courant and the $V_{P/H}$-Courant $\sigma$-models, but it's not necessarily useful).

\end{document}